\documentclass{elsart}
\usepackage[english]{babel}
\usepackage{epsfig}
\usepackage{psfrag}
\usepackage{graphicx}
\usepackage{amsmath}
\usepackage{amssymb}
\usepackage{dsfont}

\begin{document}

\begin{frontmatter}
\title{Extraction of Spectral Functions from Dyson-Schwinger Studies via the
  Maximum Entropy Method}
\author{Dominik~Nickel}
\ead{dominik.nickel@physik.tu-darmstadt.de}
\address{Institut f\"ur Kernphysik, Technische Universit\"at Darmstadt,
  D-64289 Darmstadt, Germany}

\begin{abstract}
It is shown how to apply the Maximum Entropy Method (MEM) to numerical
Dyson-Schwinger studies for the extraction of spectral functions of
correlators from their corresponding Euclidean propagators.
Differences to the application in lattice QCD are emphasized and, as an
example, the spectral functions of massless quarks in cold and dense matter are
presented.
\end{abstract}

\end{frontmatter}

\section{Introduction}
For the investigation of QCD in the strongly coupled regime, 
non-perturbative numerical methods such as lattice QCD and truncated
Dyson-Schwinger equations can be employed. Similar to lattice QCD, numerical
Dyson-Schwinger studies of QCD n-point functions are usually performed in
Euclidean space (for recent reviews
see~\cite{Alkofer:2000wg,Roberts:2000aa,Fischer:2006ub}).
Although Dyson-Schwinger studies rely on truncation schemes, they have the
great advantage, that they can in principle be solved in the continuum limit
and with much higher numerical accuracy.

An extraction of dynamical properties, in particular the spectral functions
of propagators, is however highly desirable. In quantum Monte Carlo
simulations, the Maximum Entropy Method (MEM) turns out to be an especially
suited tool and has been successfully applied in condensed
matter physics (see~\cite{Jarrell} for a review), lattice QCD in the vacuum
(see~\cite{Asakawa:2000tr} for a review), as well as at finite
temperatures~\cite{Karsch:2001uw}.
In this work we show, that it can also successfully be employed for the
extraction of spectral functions in Dyson-Schwinger studies.

The starting point for the MEM is the linear relation between the spectral
function and numerically determined Euclidean correlation functions
via generalized K{\"a}llen-Lehmann representations. Since the inversion of the
latter is in general ill-posed due to the spectral properties of the linear
operator, further knowledge has to be implemented non-linearly.
This can be done from a regularization point of view leading to the
``historical maximum entropy''~\cite{Recipes} or in a Bayesian approach
leading to the ``classic maximum entropy''~\cite{Skilling,Gull} and to
``Bryan's method''~\cite{Bryan}.

The key idea in the latter case is the interpretation of the spectral function
as a probability distribution due to its special properties and a proper
consideration of the numerical error. As a result, the MEM determines the most
plausible or expected spectral function for a given Euclidean correlator with
known errors and some prior knowledge. It does not rely on a special form of
the function and should, with decreasing errors, converge towards the exact
solution.

For numerical Dyson-Schwinger studies, the method seems especially reliable,
since the calculations can usually be performed with much higher numerical
accuracy and for much more momentum points of the correlators than in lattice
QCD.

This paper is organized as follows: In section~\ref{specfunc} we collect
those properties of spectral functions for fermions and bosons, which make them
suitable for the application of the MEM. In section~\ref{MEM} we discuss the
MEM procedure itself, in particular its adaptation to Dyson-Schwinger studies
and to fermions. In section~\ref{numimpl} we sketch the numerical
implementation. After this we show in section~\ref{colddense} some results for
massless color-superconducting fermions in dense quark-gluon matter. Finally
we summarize and conclude in section~\ref{sumconc}.

\section{Spectral functions and their properties}
\label{specfunc}
Performing calculations for propagators in Euclidean space or rather within the
imaginary time formalism means the determination of Matsubara propagators.
These are related to spectral functions in Minkowski space via the generalized
K{\"a}llen-Lehmann representation (see e.g.~\cite{LeBellac}). Using the
Euclidean conventions of~\cite{Alkofer:2000wg} we have
\begin{eqnarray}
  S(\vec{p},p_{4})
  &=&
  \int_{-\infty}^{\infty}\frac{d\omega}{2\pi}
  \frac{\rho(\omega,\vec{p})}{-ip_{4}+\mu-\omega},
\end{eqnarray}
for a Dirac fermion propagator, where $p_{4}=\omega_{n}$ denotes a given
Matsubara frequency and $\mu$ is the chemical potential. The spectral function
is given by
\begin{eqnarray}
  \label{spec1}
  \rho(\omega,\vec{p})_{\alpha\beta}
  &=&
  \frac{(2\pi)^{4}}{Z(\beta)}\sum_{l,n}
  e^{-\beta (E_{l}-\mu N_{l})}(1+e^{-\beta \omega})
  \langle l|\psi_{\alpha}|n\rangle
  \langle n|\bar{\psi}_{\beta}|l\rangle
  \times
  \nonumber\\&&
  \phantom{\frac{2\pi}{Z(\beta)}\sum_{l,n}}
  \times
  \delta\left(\omega+E_{l}-E_{n}\right)
  \delta^{3}\left(\vec{p}+\vec{p}_{l}-\vec{p}_{n}\right).
\end{eqnarray}
Therefore $\rho(\omega,\vec{p})\gamma_{4}$ is hermitian and has only positive
eigenvalues in a given Hilbert space\footnote{This argument does therefore not
  hold e.g. for gauge fields.}. Furthermore it has to fulfill the sum rule
\begin{eqnarray}
  \gamma_{4} &=&
  \frac{Z_{2}}{2\pi}\int_{-\infty}^{\infty}\!\!\!d\omega\,\,\,
  \rho(\omega,\vec{p}),
\end{eqnarray}
as a consequence of the (anti-)commutation relations. Thus
$\rho(\omega,\vec{p})\gamma_{4}$/4 can be identified with a probability
distribution, which is the key property for motivating the use of the MEM.

For a massive relativistic fermion in an isotropic, even parity and
$T$-symmetric phase, we can parameterize
\begin{eqnarray}
  \label{massiverho}
  \rho(\omega,\vec{p})
  &=&
  2\pi
  \left(
    \omega\gamma_{4}\,\rho_{e}(\omega,\vec{p})
    -i\vec{p}\cdot\vec{\gamma}\,\rho_{v}(\omega,\vec{p})
    +\rho_{s}(\omega,\vec{p})
  \right)
\end{eqnarray}
and knowing that all eigenvalues have to be positive, we get
\begin{eqnarray}
  \label{posrhoe}
  \omega\rho_{e}(\omega,\vec{p}) \geq
  \sqrt{\vec{p}^{2}\rho_{v}(\omega,\vec{p})^{2}+\rho_{s}(\omega,\vec{p})^{2}}
  \geq 0
\end{eqnarray}
and furthermore the sum rules
\begin{eqnarray}
  \label{sumrules}
  1 &=&
  Z_{2}\int_{-\infty}^{\infty}\!\!\!d\omega\,\,\,
  \omega\,\rho_{e}(\omega,\vec{p}),
  \nonumber\\
  0 &=&
  \phantom{Z_{2}}\int_{-\infty}^{\infty}\!\!\!d\omega\,\,\,
  \rho_{v}(\omega,\vec{p}),
  \nonumber\\
  0 &=&
  \phantom{Z_{2}}\int_{-\infty}^{\infty}\!\!\!d\omega\,\,\,
  \rho_{s}(\omega,\vec{p}).
\end{eqnarray}

For the application discussed in section~\ref{colddense}, we will for
simplicity restrict ourself to the chiral limit, i.e. massless fermions (the
case of massive fermions is discussed at the end of section 3). In
this case, we can rewrite the propagators by using the energy projectors
$\Lambda^{\pm}=\frac{1}{2}\left(1\pm i\gamma_{4}\frac{\vec{p}\cdot\vec{\gamma}}{\vert\vec{p}\vert}\right)$
and obtain
\begin{eqnarray}
  \rho(\omega,\vec{p})
  &=&
  2\pi
  \left(
    \rho^{+}(\omega,\vec{p})\Lambda^{+}\gamma_{4}
    +
    \rho^{-}(\omega,\vec{p})\Lambda^{-}\gamma_{4}
  \right).
\end{eqnarray}
The spectral functions $\rho^{\pm}$ then fulfill
\begin{eqnarray}
  \rho^{\pm}(\omega,\vec{p}) &>& 0,
  \nonumber\\
  Z_{2}\int_{-\infty}^{\infty}\!\!\!d\omega\,\,\,\rho^{\pm}(\omega,\vec{p})
  &=& 1,
\end{eqnarray}
with
\begin{eqnarray}
  \label{specreppm}
  S^{\pm}(\vec{p},p_{4})&=&
  \int_{-\infty}^{\infty}\!\!\!d\omega\,\,\,
  \frac{\rho^{\pm}(\omega,\vec{p})}{-ip_{4}+\mu-\omega}.
\end{eqnarray}

For completeness we also wish to show, how the solutions of the inhomogeneous
Bethe-Salpeter equation (BSE) can be used to determine the spectral functions
of mesons or diquarks. For a given current
$J^{a}(x)=\bar{\psi}(x)T^{a}\psi(x)$, the solution for the corresponding
Bethe-Salpeter amplitude $\Gamma^{a}(q;P)$ in momentum space (see
e.g.~\cite{Alkofer:2000wg}) determines the time-ordered product
\begin{eqnarray}
  \langle T\,J^{a}(x)\psi(y)\bar{\psi}(z)\rangle_{\beta}
  &=&
  {\int\!\!\frac{d^4P}{(2 \pi)4}}{\int\!\!\frac{d^4q}{(2 \pi)4}}
  \,\,e^{-iP(x-\frac{y+z}{2})+iq(z-y)}\times
  \nonumber\\&&
  \phantom{{\int\!\!\frac{d^4P}{(2 \pi)4}}}
  \times\,\,
  S(q+\frac{P}{2})\Gamma^{a}(q;P)S(q-\frac{P}{2}),
\end{eqnarray}
such that the current-current correlator is given by
\begin{eqnarray}
  \langle T\,J^{a}(x)J^{b}(y)\rangle_{\beta}  
  &=&
  \lim_{z\rightarrow y}
  \mathrm{Tr}
  \left(\langle T\,J^{a}(x)\psi(y)\bar{\psi}(z)\rangle_{\beta}T^{b}\right)
  \nonumber\\&=&
  {\int\!\!\frac{d^4P}{(2 \pi)4}}
  \,\,e^{-iP(x-y)}\,D_{ab}(P),
\end{eqnarray}
with
\begin{eqnarray}
  D_{ab}(P)&=&
  {\int\!\!\frac{d^4q}{(2 \pi)4}}
  \mathrm{Tr}\left(
    S(q+\frac{P}{2})\Gamma^{a}(q;P)S(q-\frac{P}{2})T^{b}
  \right).
\end{eqnarray}
Again $D_{ab}(P)$ possesses a generalized K{\"a}llen-Lehmann representation
and the spectral function has to fulfill positivity conditions.

\section{Maximum Entropy Method (MEM)}
\label{MEM}
The Maximum Entropy Method is a numerical tool for the inversion of
potentially ill-posed linear equations by the implementation of additional
information, i.e. constraints. It can be easily viewed from a standpoint of
regularization as adding some non-linear auxiliary conditions, leading to the
so-called ``historical maximum entropy''~\cite{Recipes}. This is known to
underfit the data by overestimating the effective number of degrees of
freedom, thus leading to solutions that are closer to the prior
estimate. Usually, the somewhat
converse Bayesian viewpoint is considered, since it essentially adjusts the
number of effective degrees of freedom and also allows for an error
estimation~\cite{Jarrell,Asakawa:2000tr}. We briefly review this here,
emphasizing the adaption to our problem.

Given a (numerically evaluated) Euclidean correlator, which is treated as
'data' $D$, the objective is to determine the most plausible (related to
``classic maximum entropy''~\cite{Skilling,Gull}) or the most expected
(related to ``Bryan's method''~\cite{Bryan}) spectral function
$\rho_{\mathrm{MEM}}$ by
taking into account prior knowledge $H(m)$ of the solution, regulated by the
prior estimate $m$ to be defined below. The key entity is the plausibility
functional $P[\rho|DH(m)]$ for the spectral function $\rho$ under given $D$
and $H(m)$.
With help of the so called ``Bayesian theorem'' for conditional plausibilities
\begin{align}
  P[XY] = P[X|Y]P[Y] = P[Y|X]P[X],
\end{align}
this can be brought into the form
\begin{eqnarray}
  \label{Prho}
  P[\rho|DH(m)] &=& \frac{P[D|\rho H(m)]P[\rho|H(m)]}{P[D|H(m)]}
  \nonumber\\&\propto&
  P[D|\rho H(m)]P[\rho|H(m)].
\end{eqnarray}
Here, we have introduced the ``likelihood function'' $P[D|\rho H(m)]$ for the
plausibility of the data $D$ under given $\rho$ and $H(m)$ and the ``prior
probability'' $P[\rho|H(m)]$ for the plausibility of $\rho$ under the prior
knowledge $H(m)$. The constant plausibility $P[D|H(m)]$ of the data $D$ under
the prior knowledge $H(m)$ can be dropped, since we normalize the plausibility
functional at the end.

Considering the function (or sequence at finite temperature)  $D$ in an
interval $[a,b]$
as uncorrelated data points obeying a Gaussian distribution
functional\footnote{
The justification of this will be discussed for a given application.}, we
have
\begin{eqnarray}
  P[D|\rho H(m)] &=& \exp\left(-L\left[\rho\right]\right),
\end{eqnarray}
with the likelihood
\begin{eqnarray}
  \label{likeL}
  L\left[\rho\right]
  &=&
  \frac{1}{b-a}\,\int_{a}^{b}\!dp_{4}
  \frac{|D(p_{4})-D[\rho](p_{4})|^{2}}{2\sigma(p_{4})^{2}}
  \nonumber\\&\simeq&
  \frac{1}{b-a}
  \sum_{i}\Delta p_{4,i}
  \frac{|D_{i}-D[\rho]_{i}|^{2}}{2\sigma_{i}^{2}},
\end{eqnarray}
where $D[\rho]$ is given by the generalized K{\"a}llen-Lehmann representation
and a measure $\mathcal{D}D$ for the discretized integral
\begin{eqnarray}
  \mathcal{D}D
  &=&
  \prod_{i}
  \sqrt{\frac{\Delta p_{4,i}}{2\pi(b-a)\sigma_{i}^{2}}}\,dD_{i}\,\,\,.
\end{eqnarray}

The prior probability $P[\rho|H(m)]$ is usually somewhat arbitrary and
essentially implements the positivity conditions non-linearly, at least from
the regularization point of view. It can be motivated by the law of large
numbers or axiomatically constructed (see
e.g.\cite{Jarrell,Asakawa:2000tr}). The key idea in the latter case is to
consider the spectral function as a probability distribution and derive the
most general functional, fulfilling the requirements of subset independence,
coordinate invariance, system independence and scaling. It is then of the form
\begin{eqnarray}
  P[\rho|H(m)] &=& \int_{0}^{\infty}\!\!\! d\alpha\,\,\,
  P[\rho|H(\alpha m)]P[\alpha|H(m)],
\end{eqnarray}
with the scaling factor $\alpha$ for the prior estimate $m$ and the
plausibility for the scaled prior estimate
\begin{eqnarray}
  P[\rho|H(\alpha m)] &=&
  \exp\left(\alpha S\left[\rho\right]\right),
\end{eqnarray}
where $S\left[\rho\right]$ is the negative semi-definite entropy
\begin{eqnarray}
  S\left[\rho\right] &=& \int_{-\infty}^{\infty}\!\!\!d\omega\,\,\,
  \left(
    \rho(\omega)-m(\omega)
    -\rho(\omega)\ln\left(\frac{\rho(\omega)}{m(\omega)}\right)
  \right)
  \nonumber\\&\simeq&
  \sum_{i} \Delta \omega_{i}
  \left(
    \rho_{i}-m_{i}
    -\rho_{i}\ln\left(\frac{\rho_{i}}{m_{i}}\right)
  \right)
  \nonumber\\&=&
  -\sum_{i}
  2\Delta \omega_{i}
  \left(\sqrt{\rho_{i}}-\sqrt{m_{i}}\right)^{2}
  +O\left((\sqrt{\rho_{i}}-\sqrt{m_{i}})^{3}\right)
\end{eqnarray}
with the measure
\begin{eqnarray}
  \mathcal{D}\rho_{\alpha}
  &\simeq&
  \prod_{i}d\sqrt{\frac{2\alpha\Delta\omega_{i}\rho_{i}}{\pi}}
\end{eqnarray}
in the saddle point approximation around the prior estimate. The scaling
factor $\alpha$ basically scales the maximum of the entropy and will be
balanced by the likelihood. Furthermore we will assume, that the plausibility
$P[{\alpha}|H(m)]$ can be dropped, which is called Laplace's rule and can be
justified a posteriori, as discussed in our example in
section~\ref{colddense}.

From Eq.~(\ref{Prho}) we therefore finally get 
\begin{eqnarray}
  P[\rho|DH(m)]
  &=&
  \frac{1}{Z}\exp\left(Q\left[\rho\right]\right),
\end{eqnarray}
with the negative semi-definite functional $Q\left[\rho\right]=\alpha
S\left[\rho\right]-L\left[\rho\right]$, $Z$ determined by normalization and
the measure
\begin{eqnarray}
  \mathcal{D}\rho
  &\simeq&
  d\alpha\,\prod_{i}d\sqrt{\frac{2\alpha\Delta\omega_{i}\rho_{i}}{\pi}}.
\end{eqnarray}

It is worth noting, that the ``historic maximum entropy''~\cite{Recipes} simply
determines the maximum, which is unique if it
exists~\cite{Asakawa:2000tr}, of the functional
$Q\left[\rho\right]$ with $\alpha$ chosen, such that $L=1$. We will
however follow ``Bryan's method''~\cite{Bryan}, aiming at the most expected
spectral function, by computing
\begin{eqnarray}
  \rho_{\mathrm{MEM}}
  &=&
  \int\!\!\mathcal{D}\rho\,\,\rho\,\, P[\rho|DH(m)]
  \nonumber\\&\simeq&
  \int_{0}^{\infty}\!\!\! d\alpha\,\rho_{\alpha}
  \,\,\,\frac{1}{Z}\int\!\!\mathcal{D}\rho\,
  \, \exp\left(Q\left[\rho\right]\right),
\end{eqnarray}
where it is assumed that $P[\rho|DH(m)]$ is sharply peaked around its maximum
$\rho_{\alpha}$. We therefore define
\begin{eqnarray}
  P[\alpha|DH(m)]
  &=&
  \frac{1}{Z}
  \int\!\!\mathcal{D}\rho\,\,\exp\left(Q\left[\rho\right]\right)
  \nonumber\\&\simeq&
  \frac{1}{Z}
  \exp\left(
    Q\left[\rho_{\alpha}\right]+
    \frac{1}{2}\sum_{k}\ln
    \left(\frac{\alpha\Delta\omega_{k}}{\lambda_{k}}\right)
  \right),
\end{eqnarray}
with $\{\lambda_{k}\}$ being the eigenvalues of
\begin{eqnarray}
  M_{ij} &=&
  \alpha \Delta\omega_{i}\delta_{ij}+
  \sqrt{\rho_i}\,\frac{\partial^{2}L}{\partial\rho_i\partial\rho_j}\,
  \sqrt{\rho_j}
\end{eqnarray}
and finally get the most expected spectral function via
\begin{eqnarray}
  \rho_{\mathrm{MEM}}
  &=&
  \int_{0}^{\infty}\!\!\! d\alpha\,\rho_{\alpha}\,\,P[\alpha|DH(m)].
\end{eqnarray}
It should be noted, that $P[\alpha|DH(m)]$ is formally not integrable due to
the saddle point approximation. However, this becomes only relevant for very
large values of $\alpha$. In any numerically considered interval the function
decreases exponentially for precise enough and many data points. For
consistency, the choice for the upper cutoff for $\alpha$
should always be quoted. In the ``classical maximum entropy'', the most
plausible spectral function $\rho_{\hat{\alpha}}$ is determined by maximizing
$Q\left[\rho\right]$ and $P[\alpha|DH(m)]$ simultaneously. In our case, due to
(in principle) arbitrarily many data points, this turns out and is
known~\cite{Jarrell} to agree with the most expected spectral function
$\rho_{\mathrm{MEM}}$.

In comparison to previous applications, we have formulated the MEM for
arbitrarily discretized functions $D$ and $\rho_{\alpha}$, since we want to
deal with (in principle) continuous functions from truncated Dyson-Schwinger
calculations. Therefore,
the single-value decomposition as proposed in the Bryan algorithm~\cite{Bryan}
for the numerical determination of $\rho_{\alpha}$ does not work. However, our
new treatment of the spectral function opens the possibility
of a better suited discretization, which can be adopted to the specific form
of $\rho_{\alpha}$. In this way, we are also able to significantly reduce the
number of points, which are needed for the numerically discretized spectral
function.

As already mentioned, the Bayesian approach offers the possibility of an error
estimation. If we consider an interval $I=[\omega_{1},\omega_{2}]$ of the
spectral function, the expectation value of $\rho_{a}$ for fixed $\alpha$ in
this interval for constant weighting is given by
\begin{eqnarray}
  \langle\rho_{\alpha}\rangle_{I}
  &=&
  \frac{1}{\omega_{2}-\omega_{1}}\int_{I}d\omega\,\,\rho_{\alpha}(\omega )
\end{eqnarray}
and, hence, for the most plausible function by
\begin{eqnarray}
  \langle\rho_{\mathrm{MEM}}\rangle_{I}
  &=&
  \int_{0}^{\infty}\!\!\!
  d\alpha\,\langle\rho_{\alpha}\rangle_{I}P[\alpha|DH(m)].
\end{eqnarray}
For the variance around this value, we first consider~\cite{Skilling,Gull}
\begin{eqnarray}
  \langle\delta\rho_{\alpha}(\omega_{i})\delta\rho_{\alpha}(\omega_{j})\rangle
  &\simeq&
  4
  \sqrt{\rho_{\alpha}(\omega_{i})\rho_{\alpha}(\omega_{j})}
  \left\langle
  \delta\sqrt{\rho_{\alpha}(\omega_{\phantom{j}\!\!i})}
  \delta\sqrt{\rho_{\alpha}(\omega_{j})}
  \right\rangle
  \nonumber\\&\simeq&
  -\sqrt{\rho_{\alpha}(\omega_{i})\rho_{\alpha}(\omega_{j})}
  \left(M^{-1}\right)_{ij}
  \nonumber\\&\simeq&
  -\left.
    \frac{\partial^{2}Q}{\partial\rho_i\partial\rho_j}
  \right|_{\rho=\rho_{\alpha}},
\end{eqnarray}
with fixed $\alpha$ and
$\delta\rho_{\alpha}(\omega)=\rho(\omega)-\rho_{\alpha}(\omega)$. Therefore in
the given interval $I$, we get
\begin{eqnarray}
  \left\langle \left(\delta \rho_{\alpha}\right)^{2}\right\rangle_{I}
  &\simeq&
  -\frac{1}{(\omega_{2}-\omega_{1})^{2}}
  \int_{I\times I} d \omega d \omega'\,
  \left.\left(
      \frac{\delta^{2}Q}{\delta\rho(\omega)\delta\rho(\omega')}
    \right)^{-1}\right|_{\rho=\rho_{\alpha}}
\end{eqnarray}
and
\begin{eqnarray}
  \langle(\delta\rho_{\mathrm{MEM}})^{2}\rangle_{I}
  &=&
  \int_{0}^{\infty}\!\!\! d\alpha\,
  \left\langle \left(\delta\rho_{\alpha}\right)^{2}\right\rangle_{I}
  P[\alpha|DH(m)].
\end{eqnarray}

At the end of this section, we want to indicate, how the MEM can be applied for
massive fermions. For $\rho_{e}(\omega,\vec{p})$, the upper procedure can be
performed analogously due to the properties given in Eqs.~(\ref{posrhoe}) and
Eqs.(\ref{sumrules}). On the other hand, we would need to extend the procedure
for the whole propagator by utilizing that $\rho(\omega,\vec{p})\gamma_{4}/4$
can be considered as a probability distribution. With the propagator again
denoted as data $D$, we generalize
\begin{eqnarray}
  L\left[\rho\right]
  &\rightarrow&
  \frac{1}{b-a}\,\int_{a}^{b}\!dp_{4}
  \frac{
    \mathrm{Tr}\left((D(p_{4})-D[\rho](p_{4}))^{\dagger}
      (D(p_{4})-D[\rho](p_{4}))\right)
  }{2\sigma(p_{4})^{2}},
  \nonumber\\
  S\left[\rho\right]
  &\rightarrow&
  \int_{-\infty}^{\infty}\!\!\!d\omega\,\,\,
  \mathrm{Tr}\left(\left(
    \rho(\omega)-m(\omega)
    -\ln\left(\rho(\omega)m(\omega)^{-1}\right)\rho(\omega)
  \right)\gamma_{4}\right),
\end{eqnarray}
where the prior estimate $m$ is now matrix-valued. Since
$\rho(\omega,\vec{p})\gamma_{4}/4$ is hermitian and positive, it can be
written as $g^{\dagger}\rho_{D}g$, with $g\in \mathrm{U(4)}$ and $\rho_{D}$ a
diagonal matrix with positive eigenvalues. The unconstrained measure
$\mathcal{D}\rho$ then becomes 
\begin{eqnarray}
  \mathcal{D}\rho
  &\rightarrow&
  d\alpha\,\prod_{i,a}d\sqrt{\frac{2\alpha\Delta\omega_{i}\rho_{i,a}}{\pi}}
  \,\,d\lambda_{i},
\end{eqnarray}
where $\rho_{i,a}$ is the $a$-th eigenvalue and $d\lambda_{i}$ the Haar measure
for the group $\mathrm{U(4)}$ at an energy $\omega_{i}$. However, the path
integral is usually constrained by symmetries, i.e.
$\rho(\omega,\vec{p})\gamma_{4}=h^{\dagger}\rho(\omega,\vec{p})\gamma_{4}h$
for $h\in\mathcal{H}$. It can be easily seen that the group needs to be only
integrated over the factor group of $\mathrm{U(4)}$ and the
conjugate closure of $\mathcal{H}$ and that only the independent eigenvalues
need to be considered. For the case given by Eq.~(\ref{massiverho}), we obtain
with
\begin{eqnarray}
  \rho(\omega,\vec{p})
  &=&
  \rho^{+}(\omega,\vec{p})\,\Lambda_{\omega,\vec{p}}^{+}\,\gamma_{4}+
  \rho^{-}(\omega,\vec{p})\,\Lambda_{\omega,\vec{p}}^{-}\,\gamma_{4}\,\,,
\end{eqnarray}
where $\Lambda_{\omega,\vec{p}}^{\pm}=\frac{1}{2}
(1\pm(i\cos\theta(\omega,\vec{p})\gamma_{4}\frac{\vec{p}\cdot\vec{\gamma}}{\vert\vec{p}\vert}-
\sin\theta(\omega,\vec{p}) \gamma_{4}))$, simply
\begin{eqnarray}
  \mathcal{D}\rho
  &\rightarrow&
  d\alpha\,\prod_{i}
  d\sqrt{\frac{2\alpha\Delta\omega_{i}\rho^{+}_{i}}{\pi}}\,\,
  d\sqrt{\frac{2\alpha\Delta\omega_{i}\rho^{-}_{i}}{\pi}}\,\,
  \frac{d\theta_{i}}{\pi}.
\end{eqnarray}
In the approximation of setting the integrand of the $\theta$-integration to
be constant and equal to the value $\theta_{i}$ given by the maximization of
the functional $Q[\rho]$, the practical MEM procedure again becomes similar to
the upper case.

\section{Numerical implementation}
\label{numimpl}
For the numerical determination of the most plausible spectral function, we
first need the data $D$ with errors $\sigma$ on an interval $[a,b]$.
Furthermore we have to choose a suitable interval $[\omega_{1},\omega_{2}]$
for the considered part of the spectral function and a prior estimate $m$.
The interval $[\omega_{1},\omega_{2}]$ is usually suggested by the involved
scales and can be chosen to be rather large. The prior estimate $m$ is in our
case taken as a constant and can be estimated from the sum rules
(Eqs.~(\ref{sumrules})), if the spectral function varies only
in a certain interval $\Delta I$. It can also be adopted to the knowledge from
other methods. Since the main purpose is the implementation of positivity, the
results turn out to be comparatively insensitive to its choice. The advantage
of our method is that, after an eventual test calculation, we can choose the
discretization, i.e. the abscissas and weights, of the spectral function, such
that it is interpolated by a small number of points.

We summarize the procedure as follows:
\begin{itemize}
\item Take the data $D$ with errors $\sigma$ on an interval $[a,b]$ and choose
  a prior estimate $m$ on an interval $[\omega_{1},\omega_{2}]$.
\item Determine the maximum of the functional $Q[\rho]=L[\rho]-\alpha
  S[\rho]$ for a fixed value of $\alpha$. Due to its simple form, the
  Marquardt-Levenberg method~\cite{Recipes} is very well suited.
\item Choose a discretized interval for $\alpha$, such that $P[\alpha|DH(m)]$
  is strongly peaked. $Z$ is determined by normalization.
  For consistency, the choice for the upper cutoff for $\alpha$ should always
  be quoted.
\item Calculate $\rho_{\mathrm{MEM}}$ and eventually
  $\langle\rho_{\mathrm{MEM}}\rangle_{I}$ and
  $\langle(\delta\rho_{\mathrm{MEM}})^{2}\rangle_{I}$ for a chosen interval
  $I$.
\end{itemize}

\section{Spectral functions of quarks in cold dense matter}
\label{colddense}
\subsection{Color-superconducting quark matter}
As an example, we now want to present results for spectral functions of
massless quarks in dense matter at vanishing temperature, as they have been
determined in~\cite{Nickel:2006vf}. We consider the gapped channel in the
color-superconducting $2SC$ phase at a quark chemical potential of
$\mu=1\mathrm{GeV}$. The propagator is then of the form
\begin{eqnarray}
  S(p_{4},\vec{p}) &=&
  S^{+}(p_{4},\vec{p})\Lambda_{\vec{p}}^{+}\gamma_{4}+
  S^{-}(p_{4},\vec{p})\Lambda_{\vec{p}}^{-}\gamma_{4}
\end{eqnarray}
and $S^{+}(p_{4},\vec{p})$ is related to $\rho^{+}(\omega,\vec{p})$ by
Eq.~(\ref{specreppm}). For the bare normal quark propagator with a constant
gap $\Delta$, the spectral function is then given by
\begin{eqnarray}
  \label{rhoBCS}
  \rho^{+}(\omega,\vec{p}) &=&
  \phantom{+}\left(\frac{1}{2}+\frac{\mu-p}{2E_{\Delta}}\right)
  \delta\left(\omega+E_{\Delta}-\mu\right)
  \nonumber\\&&
  +\left(\frac{1}{2}-\frac{\mu-p}{2E_{\Delta}}\right)
  \delta\left(\omega-E_{\Delta}-\mu\right),
\end{eqnarray}
with $p=|\vec{p}|$ and
$E_{\Delta}=\sqrt{\left(p-\mu\right)^{2}+\Delta^{2}}$. We will see below, how
a non-trivial $p_{4}$-dependence generates a finite width.

\subsection{Input data and error estimate}
As described in section~\ref{MEM}, the main input for the MEM is the data with
a proper error estimate. In Dyson-Schwinger studies, these are obtained by
self-consistent solutions of truncated integral equations. To lowest order, the
error of $S^{+}$ therefore scales with the error of the numerical integrals,
which determine the normal and anomal self energies $\Sigma^{+}$ and
$\Phi^{+}$
(see~\cite{Nickel:2006vf} for details). In our case, we have chosen a simple
Riemann quadrature for the multidimensional integrals, due to the
principle-value-type behavior around the Fermi surface for ungapped channels.
The error is therefore of order $O(h)$, where $h$ is a scaling factor of the
integration mesh. For the error estimation, we therefore calculate the
propagator for two different $h$ and extrapolate linearly to $h=0$. The data
are then taken as the result for the smaller scaling factor $h$ and the
error as the difference between these data and the extrapolated
result. In addition, the errors around nearest neighbors are averaged in
order to avoid (artifically appearing) vanishing errors.

We also have to justify, that correlations between the data points are
negligible for the likelihood in the form given in Eq.~(\ref{likeL}). Since
our numerical integrals for different values of $p_{4}$ are in principle
independent, this is assumed to be true, at least when the discretized data is
coarser than the numerical integral of the self energies.

\begin{figure}
{\hspace{0.cm}\includegraphics[width=5.5cm,angle=-90]{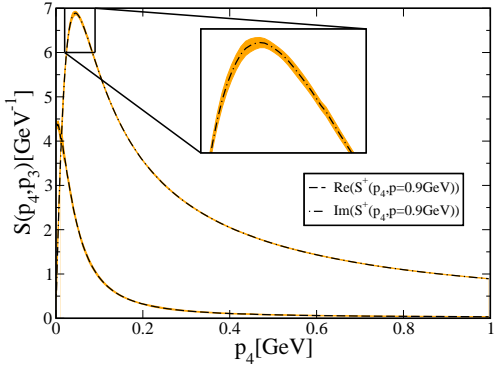}}
\caption{Real and imaginary part of the quasiparticle propagator $S^{+}$ in
  the gapped channel of the 
  $2SC$ phase at a chemical potential of $\mu=1\mathrm{GeV}$ and momentum
  $p=0.9\mathrm{GeV}$ as a function of the Euclidean energy $p_{4}$.
  The errors are given as shaded regions around the lines and are of the order
  of their thickness.
}
\label{dataSplus}
\end{figure}

The input data for $S^{+}(p_{4},\vec{p})$ for the following example is 
chosen on an interval $[0\mathrm{GeV},1\mathrm{GeV}]$ for the gapped channel
in the $2SC$ phase at a quark chemical potential of $\mu=1\mathrm{GeV}$. As
an illustration, the input for momentum $p=0.9\mathrm{GeV}$ is shown in
Fig.~\ref{dataSplus}. We consider the input as continuous due to our many data
points and it has small errors of less than $1\%$ in absolute value
above $30\mathrm{MeV}$.
 
\subsection{Choice of the prior estimate}
\begin{figure}
{\hspace{0.cm}\includegraphics[width=5.5cm,angle=-90]{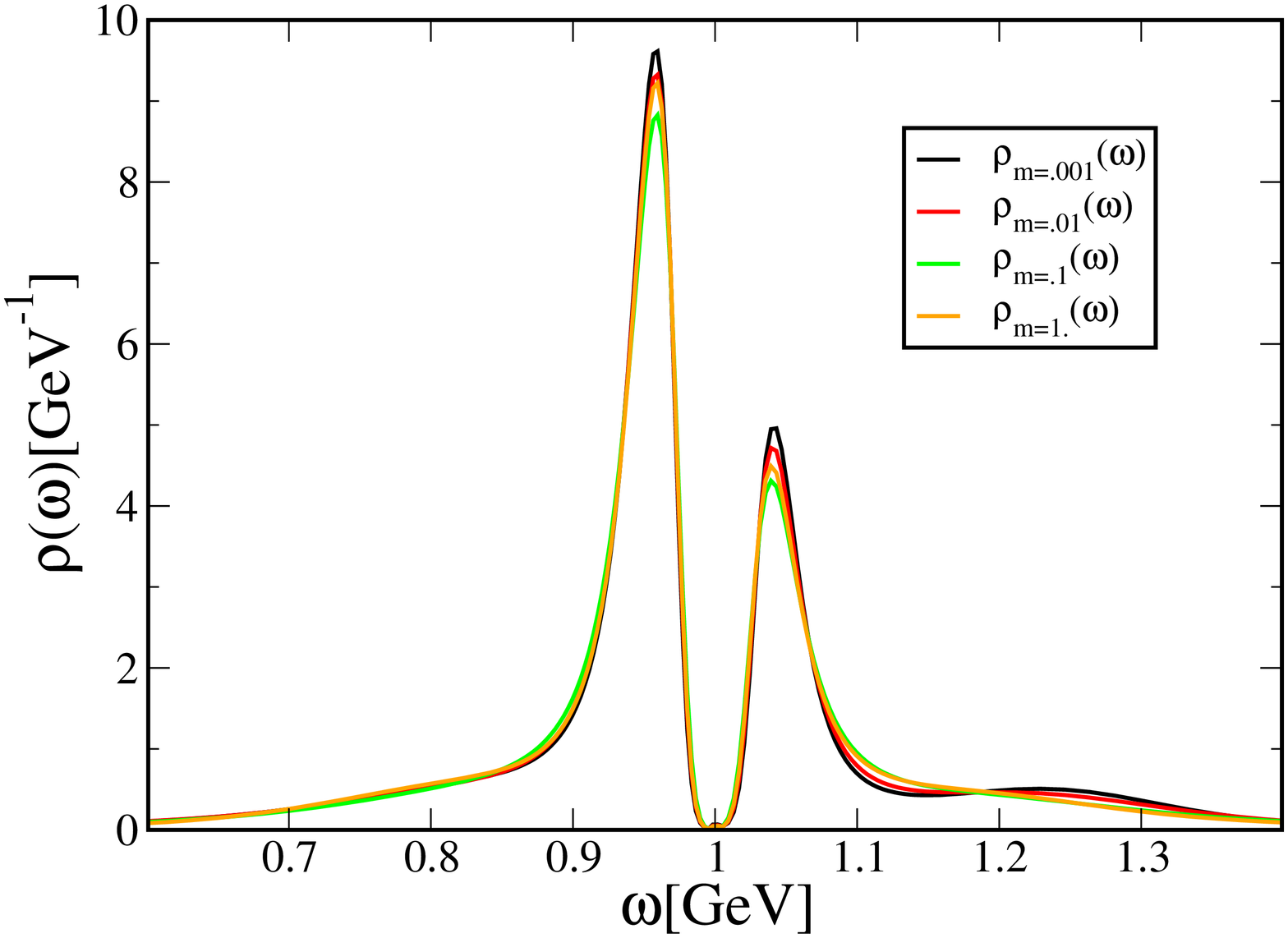}}
{\hspace{0.cm}\includegraphics[width=5.5cm,angle=-90]{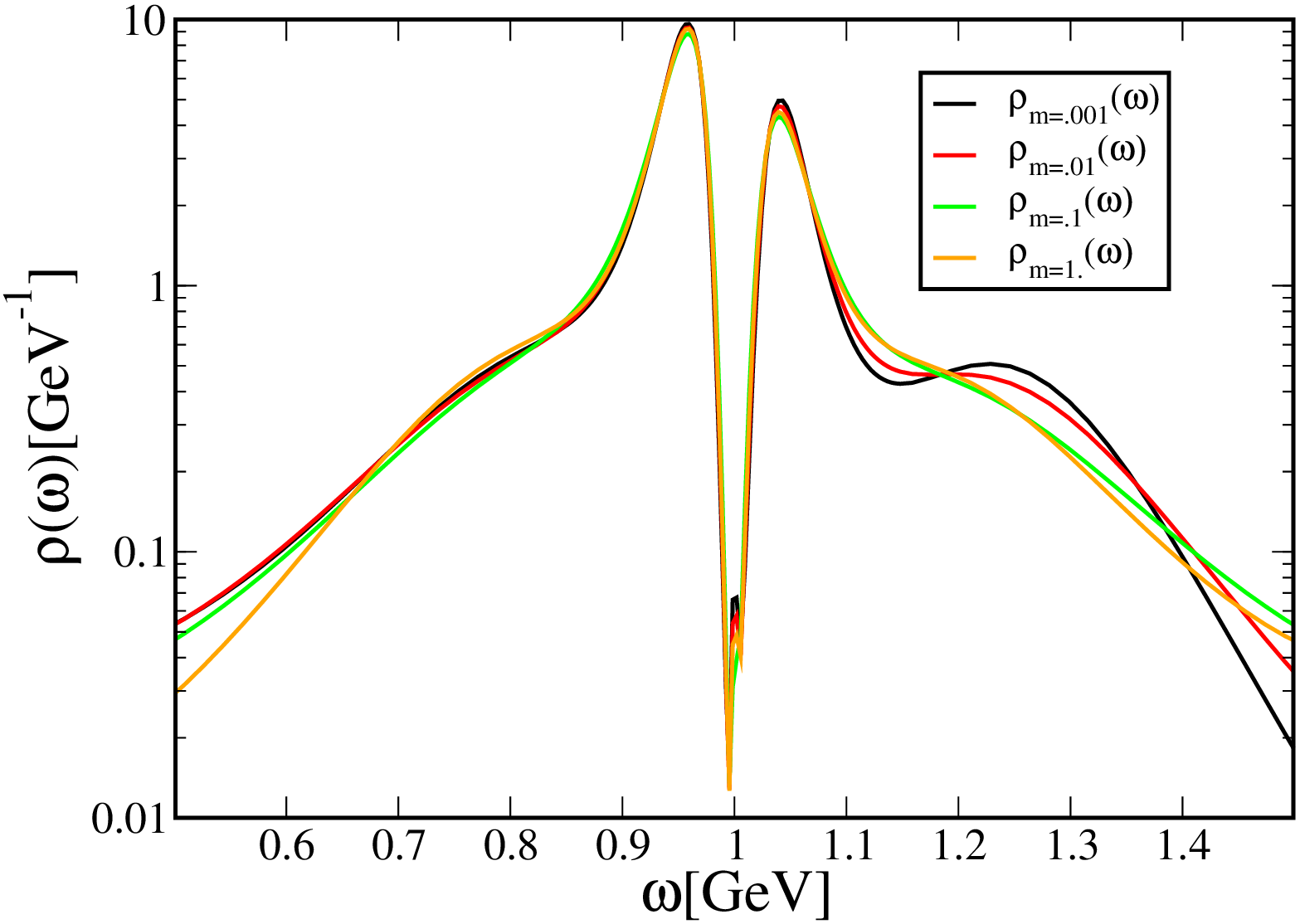}}
\caption{The most expected spectral function $\rho^{+}(\omega)$ in the gapped
  channel of the $2SC$ phase at momentum $p=0.9\mathrm{GeV}$ for constant prior
  estimates $m=0.001,0.01,0.1$ and $1.0\,\mathrm{GeV}^{-1}$ in linear (left)
  and logarithmic (right) presentation. 
}
\label{varym}
\end{figure}

For the data input with given errors, we now need to choose an interval
for the spectral function and a non-vanishing prior estimate. We take
the comparatively large interval
$[-1.5\mathrm{GeV},2\mathrm{GeV}]$. Furthermore we choose an interval for
$P[\alpha|DH(m)]$ as discussed in the following subsection.
For different prior estimates $m=0.001,0.01,0.1$ and $1.0\,\mathrm{GeV}^{-1}$
at momentum $p=0.9\mathrm{GeV}$, we then obtain the most expected spectral
functions as shown in Fig.~\ref{varym}.
It turns out, that the extracted spectral function is remarkably insensitive
on the variation of the prior estimate, even when varying it by more than three
orders of magnitude. This is mainly related to the small errors of our data.
We therefore fix $m=0.1\mathrm{GeV}^{-1}$ in the following.

\subsection{The $\alpha$-dependence}
It is also interesting to investigate the $\alpha$-dependence of the maximum
of the functional $Q[\rho]$ as well as of $P[\alpha|DH(m)]$.
Since $P[\alpha|DH(m)]$ shows a pronounced maximum at $\alpha_{max}$, we
choose the interval $I=[\alpha_{low},\alpha_{high}]$ for $P[\alpha|DH(m)]$ to
be non-vanishing and normalized, such that $P[\alpha|DH(m)]>10^{-1}\times
P[\alpha_{max}|DH(m)]$ for $\alpha\in I$.

\begin{figure}
{\hspace{0.cm}\includegraphics[width=5.5cm,angle=-90]{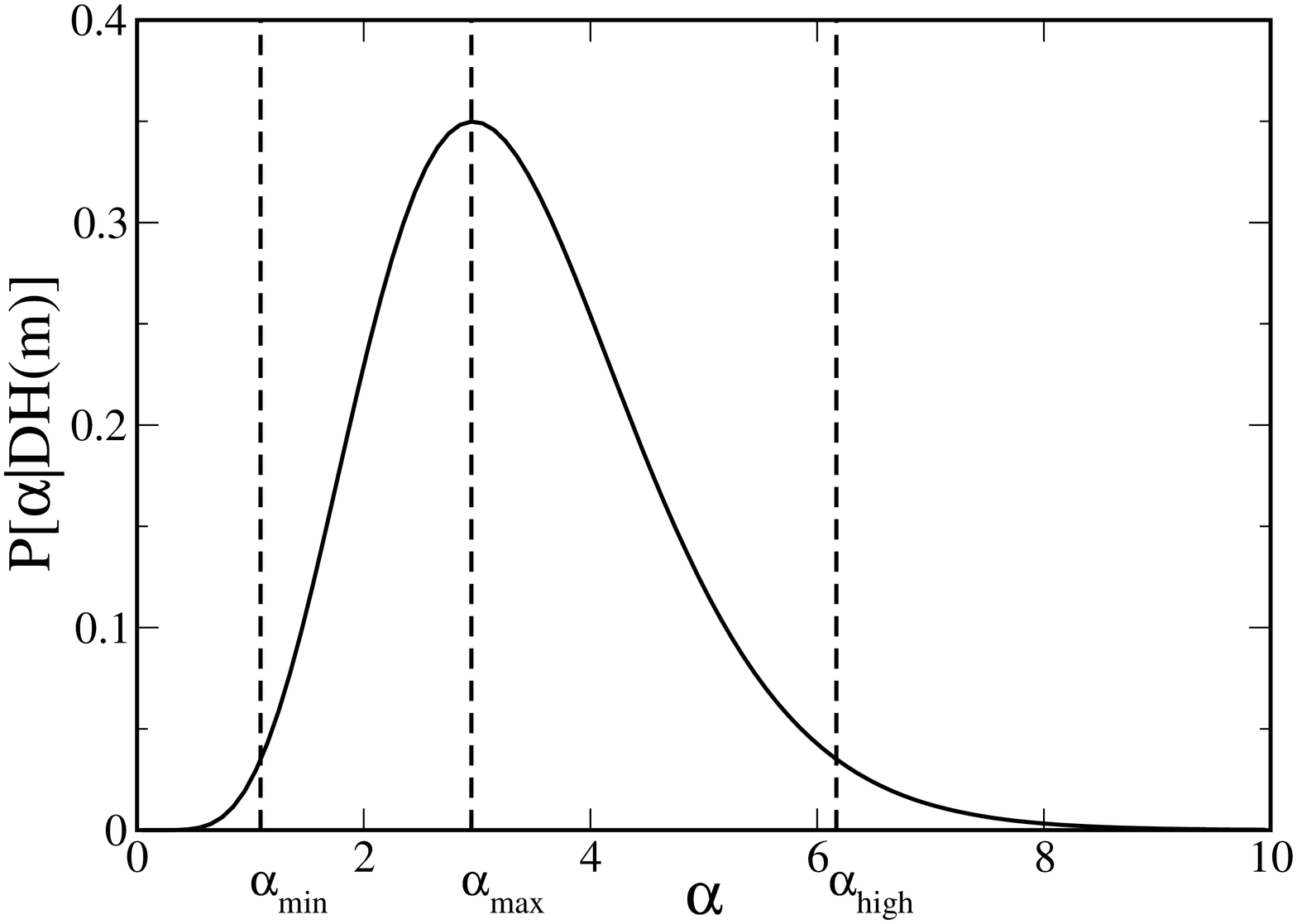}}
{\hspace{0.cm}\includegraphics[width=5.5cm,angle=-90]{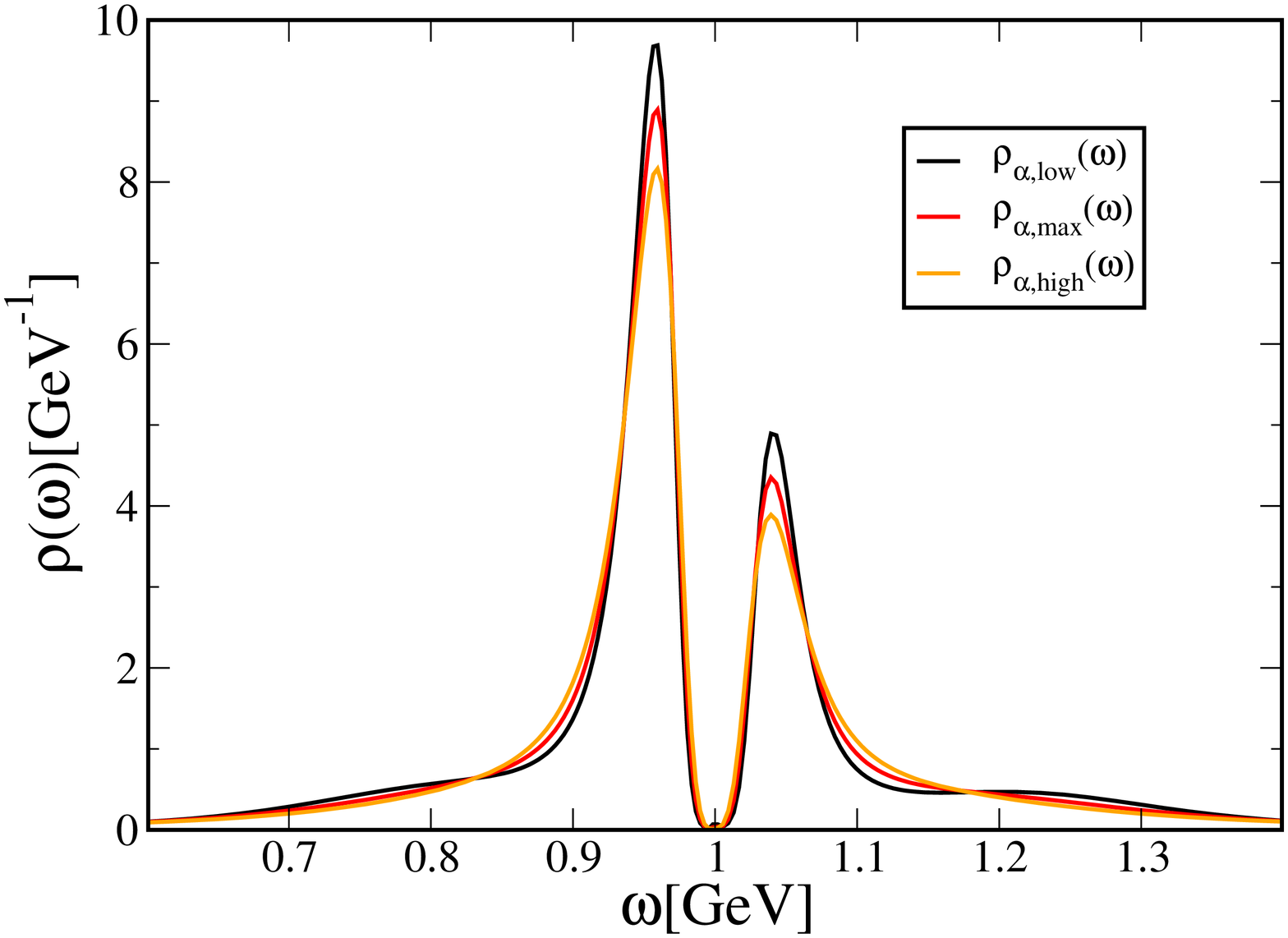}}
\caption{The function $P[\alpha|DH(m)]$ normalized between $\alpha_{min}$ and
  $\alpha_{high}$ (see text) with maximum $\alpha_{max}$ (left) and the maxima
  $\rho_{\alpha,low/max/high}$ of the functional $Q[\rho]$ for given $\alpha$
  (right). Both for the gapped
  channel of the $2SC$ phase at momentum $p=0.9\mathrm{GeV}$.
}
\label{Palpha}
\end{figure}

Again, for momentum $p=0.9\mathrm{GeV}$, the results are shown in
Fig.~\ref{Palpha}. On the left-hand side, we show $P[\alpha|DH(m)]$,
normalized on $I$. On the right-hand side, we present the maximum of $Q[\rho]$
for $\alpha=\alpha_{min},\alpha_{max}$ and $\alpha_{high}$. They are only
weakly varying, even when comparing the border of the interval $I$ to the
maximum.
This also substantiates Laplace's rule for $P[\alpha|H(m)]$ a posteriori,
assuming that its $\alpha$-dependence is weaker.

Apart from this, in Fig.~\ref{Palpha}, the most expected spectral function is
equal to the most probable spectral function given at $\alpha_{max}$ and
therefore not shown. Thus the ``classic maximum
entropy'' gives very similar results as ``Bryan's method'' in our case of
``many'' data points and small errors.

\subsection{Error estimate and final result}
\begin{figure}
\begin{picture}(10,150)
  \put(0,-2){\hspace{0.cm}\includegraphics[width=6.5cm]{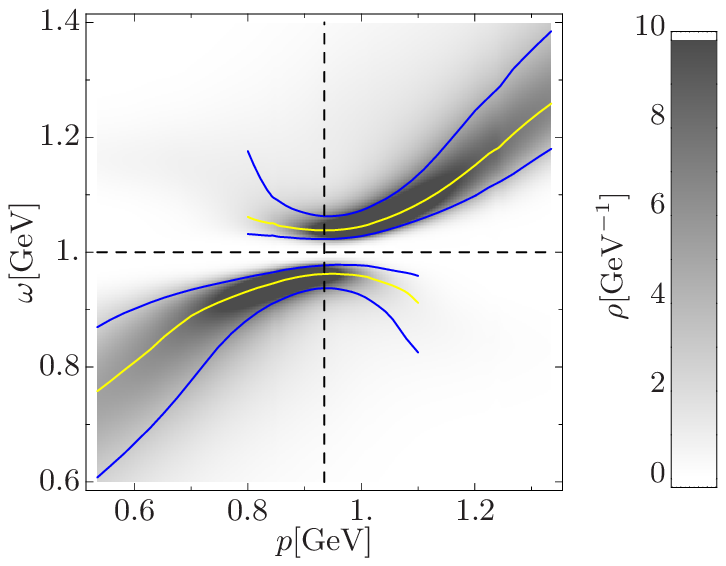}}
  \put(205,150){\hspace{0.cm}\includegraphics[width=5.5cm,angle=-90]{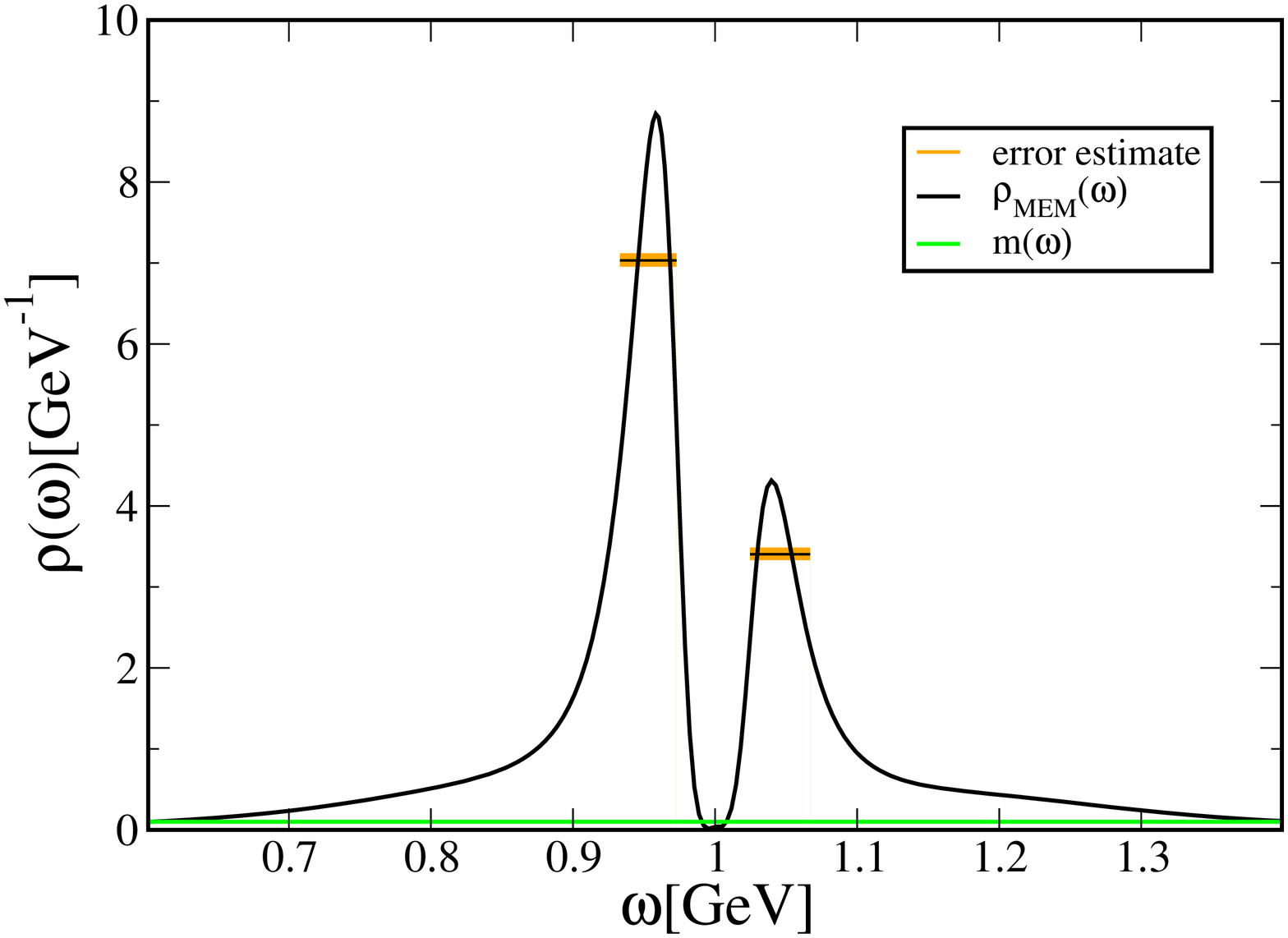}}
\end{picture}
\caption{A contour plot of most expected spectral density of the gapped $2SC$
  phase at $\mu=1\mathrm{GeV}$ as described in the text (left) and the
  spectral function for momentum $p=0.9\mathrm{GeV}$ with the expectation
  value within the FWHM and its error estimate as shaded background (right).}
\label{rho2SC}
\end{figure}

Finally we are able to perform an error estimation around the expected
value of a given interval. Since we have two pronounced peaks for
quasiparticles and quasiparticle-holes in the spectral
function, we choose the intervals associated to their full width at half
maximum (FWHM).
The result for momentum $p=0.9\mathrm{GeV}$ is shown on the right in
Fig.~\ref{rho2SC}. Again, the errors turn out to be very small.

On the left-hand side we show a contour plot of the spectral density as a
function of the energy $\omega$ and momentum $p$. The light (online yellow)
line shows the
maxima of the quasiparticle and quasiparticle-hole branches. For fixed momentum
$p$, the difference between the dark (online blue) lines below and above the
light (online yellow) line gives
the FWHM of the corresponding peak.
The latter is neglected in BCS-type spectral functions as given in
Eq.~(\ref{rhoBCS}).

\section{Summary and conclusions}
\label{sumconc}
In this paper we have outlined, how the MEM can be adapted to numerical
Dyson-Schwinger studies, in particular to fermions.
It turns out, that the extracted spectral functions are much more reliable and
stable against variation, than in applications of the MEM in lattice QCD.
Reasons for this are the comparatively small errors on the input functions
and the almost arbitrary large number of data points.
Comparing however to the systematic error, coming from necessary truncations in
Dyson-Schwinger studies, this error is negligible.
Therefore this method can be useful for further applications in mesons or
diquarks investigations, since currently all calculations have to be extended
to complex momenta (see \cite{Alkofer:2002bp,Alkofer:2003jj}).
Avoiding this and reducing the numerical effort drastically, the MEM might
help to improve or extent known truncation schemes.

\section*{Acknowledgments}
I thank Reinhard Alkofer and Jochen Wambach for helpful discussions as well
as a critical reading of the manuscript.

This work has been furthermore supported in part by the Helmholtz association
(Virtual Theory Institute VH-VI-041) and by the BMBF under grant number
06DA916.

\end{document}